# A combined criterion of surface free energy and roughness to predict the wettability of non-ideal low-energy surfaces

Majid Shaker, Erfan Salahinejad *

Faculty of Materials Science and Engineering, K. N. Toosi University of Technology, Tehran, Iran

**Abstract**

The significance of wettability between solid and liquid substances in different fields encourages scientists to develop accurate models to estimate the resultant apparent contact angles. Surface free energy (SFE), which is principally defined for ideal (flat) surfaces, is not applicable to predict the wettability of real (rough) surfaces. This paper introduces a new parameter, namely normalized surface free energy (NSFE) as a combination of SFE and roughness, to predict the contact angle of liquids on non-ideal low-energy surfaces. The remarkable consistency of the predicted and measured contact angles of liquids on some rough surfaces also confirm the validity of the approach.

---

* Corresponding Author: Email Address: <salahinejad@kntu.ac.ir>

## 1. Introduction

Each material, including solid and liquid, comprises bulk and surface with their own specific properties. The majority of interactions between substances start from their surfaces. Thus, the characteristics of surface, rather than bulk, principally determine the behavior of the material. One of the most significant factors in this regard is "solid-gas surface free energy". By tailoring the surface energy of a solid, it is possible to control reaction rate, wetting behavior, etc, which is critical for applications like catalysis, oil and gas industry, food production, physical and chemical vapor deposition, biology, wood and textile [1-9]. From this viewpoint, solid materials are classified in two groups: high- and low-surface energy materials, with surface tensions of hundreds to even thousands of mN/m and below 100 mN/m, respectively. Anyways, the determination of surface energy for solid materials mostly requires complicated calculations and/or experiments [10-14].

The most common method to determine the surface energy of solids is based on the measurement of contact angle due to its easy process. In this approach, the contact angle of two liquids on the surface of the studied substance is measured by the sessile drop method and then is used in different equations to determine surface energy. Fowkes [15-17], Owens-Wendt [18], Van Oss-Chaudhury-Good [19, 20], Fox-Zisman [21], Neumann [22] are of methods to do so. All calculations and equations are derived based on the primary assumption that the surface is ideal, i.e. free of roughness. Nonetheless, to the best of our knowledge, there is no method reported to predict the contact angle of a liquid on random shaped non-ideal (rough) surfaces. In this regard, the current paper introduces a novel approach which uses a combined concept of surface energy and roughness, to predict the contact angles of liquids on a given rough surface. Note that the Wenzel and Cassie models merely establish a correlation among the ideal contact angle, the real contact angle and a roughness-dependent factor (the area fraction of the solid surface), which is different with the concern of the current paper. It is also noteworthy to mention that this model is only applicable to low surface energy materials, due to some primary assumptions used in the development of the model, as detailed below.

## 2. Model Setup

Surface free energy (SFE) is a fundamental property of surface which is defined for ideal (flat) configuration; that is, SFE is independent upon roughness which is regarded as another characteristic of real surfaces. Indeed, most of real surfaces are heterogeneous and consist of two materials, a solid substance and air, where both of the constitutes are in contact with wetting liquids when the surface is hydrophobic. Hence, it is almost impossible to determine the SFE of a real surface. The situation is persistent even for surfaces which obey the Wenzel [23] and Cassie-Baxter [24] models used generally to characterize wettable and non-wettable surfaces, respectively. To overcome this challenge, a new surface parameter is here defined as called "Normalized Surface Free Energy (NSFE)" which combines the concept of SFE and roughness, thereby enabling us to predict the wettability of different liquids on such real surfaces.

In order to launch the arrangement of the proposed concept, the BS EN 828:2013 standard [25] is used, as below:

$$0.5\gamma_{lg}(1+\cos\theta_Y) = \sqrt{\gamma_{sg}^d \gamma_{lg}^d} + \sqrt{\gamma_{sg}^p \gamma_{lg}^p} \qquad (1)$$

where $\gamma_{lg}$ is the surface tension of a liquid, $\theta_Y$ is the Young contact angle on a solid surface, $\gamma_{sg}^d$ is the dispersive component of the solid SFE, $\gamma_{lg}^d$ is the dispersive component of the liquid SFE, $\gamma_{sg}^p$ is the polar component of the solid SFE and $\gamma_{lg}^p$ is the polar component of liquid

This is the accepted manuscript (postprint) of the following article:
M. Shaker, E. Salahinejad, *A combined criterion of surface free energy and roughness to predict the wettability of non-ideal low-energy surfaces*, Progress in Organic Coatings, 119 (2018) 123-126.
https://doi.org/10.1016/j.porgcoat.2018.02.028SFE. However, Eq. 1 is merely practical for ideal surfaces characterized to be entirely homogeneous and flat. Herein, to define NSFE as an extension of SFE, the master key is that the apparent contact angle ($\theta_a$) on a real (rough) surface is substituted for Young contact angle ($\theta_Y$) in Eq. 1, giving Eq. 2:.

$$0.5\gamma_{lg}(1+\cos\theta_\alpha) = \sqrt{\gamma_{sg}^{d*}\gamma_{lg}^{d}} + \sqrt{\gamma_{sg}^{p*}\gamma_{lg}^{p}} \qquad (2)$$

Accordingly, $\gamma_{sg}^{d*}$ and $\gamma_{sg}^{p*}$ is converted into the dispersive and polar components of NSFE on the heterogeneous surface, respectively. For a given liquid with the known liquid-gas SFE components which are clearly independent of the solid surface under consideration, there are two unknown parameters consisting of $\gamma_{sg}^{d*}$ and $\gamma_{sg}^{p*}$ in Eq. 2. Therefore, the utilization of two different liquids with the known properties yields the mentioned unknown parameters. Eventually, $\gamma_{sg}^{*}$ as the NSFE of the real solid is calculated as below:

$$\gamma_{sg}^{d*} + \gamma_{sg}^{p*} = \gamma_{sg}^{*} \qquad (3)$$

After determining the components of NSFE, in order to predict the wettability of other liquids on the same real surface, Eq. 2 should be again used. But in this occasion, there is only one unknown which is the contact angle of the third liquid. It should be noted that the application of this method has two constraints which originate from the conditions of the BS EN 828:2013 standard:

1) The primary assumption for establishing Eq. 1 is that the equilibrium vapor pressure of the liquid on the solid surface is zero, which is satisfied only for low-SFE and hydrophobic surfaces Thus, Eq. 1 and thereby Eq. 2 can be merely used for low-SFE materials.

2) Concerning the first two liquids used to determine the NSFE of the surface, one should have a dominant dispersive component of surface tension like diiodomethane and the other should present the domination of the polar component like water.

## 3. Model Verification

To ascertain the validity of the introduced approach, it is required to compare the amounts calculated and measured for several low-energy surfaces. To do so, the surface preparation and structure of four instances are first described below.

I) Nande et al. [26] deposited temperature-dependent switchable coatings of silica nanoparticles and 1H, 1H, 2H, 2H-perfluorooctyltrichlorosilane (PFOTS) on a steel substrate. In Fig. 1A, a micron-sized roughness is observed on the deposited surface. Also, in a higher-magnification micrograph (Fig. 1B), it can be seen that an overhanging-like microstructure is involved in the whole coating.

II) Deng et al. [27] coated paraffin candle soot on a glass slide with a loose, fractal-like network microstructure. The soot film consisting of carbon nanoparticles was coated with a silica shell via a chemical-vapor deposition method. Then, by annealing the coating, carbon inside the shells is combusted and the shell thickness is reduced to 20±5 nm, whereas the network structure is kept at the desired roughness.

III) Li et al. [28] deposited silica nanotubes modified with 1H,1H,2H,2H-perfluorodecyltrichlorosilane (PFDTS) on glass slides via a spray-coating method. The microstructure of the optimal surface is presented in Fig. 2, depicting the tubular structure of multi-walled carbon nanotubes used in the coating process as the template.

IV) Zhang et al. [29] developed coatings containing silicone nano-filaments modified by oxygen plasma processing and then with PFDTS onto glass slides, as illustrated in Fig. 3 The surface tension data of the liquids used to evaluate the wettability of the surfaces of the above four examples are listed in Table 1. Also, the comparison between the predicted and measured contact angles are tabulated in Table 2. It can be seen that the introduced approach predicts the contact angle of the third liquids having a very low polar surface tension component with errors below a few degrees, suggesting the validity of this model. Typically, there are two types of errors in the data: i) the difference between the predicted and measured contact angles for a given set of the three liquids, which is provided in a row of the table and ii) the difference of the NSFEs calculated by the different pairs of the two known liquids, which is presented in the different rows of the table. Albeit for a given surface, these errors can originate from the insufficient satisfaction of the model assumptions, as discussed below:

    1. The calculation of NSFE by Eq. 2 is based on the assumption that the surface under the liquid droplet is well-defined. That is, the surface can be rough in contrast to Eq. 1 used to determine SFE; however, it should be only composed of two phases, including a low-SFE solid and air. The latter phase originates from the rough character of the surface. As a delicate counterexample for this assumption, suppose that a low-SFE material is deposited on a high-SFE substrate with the aim of hydrophobization. In the meanwhile, inevitable uncoated areas of the substrate can violate the first assumption of the model where the SFE of the surface should obligatorily be low. On the other hand, a consideration to this type of error can be employed to analyze the deposition efficiency of such coating processes. That is, the convergence of the measured and calculated contact angles dictates the homogeneity and merit of the hydrophobicating process.

    2. Provided that the first two liquids (Liquids 1 and 2) are not perfectly polar (surface tension dispersive component = 0) and non-polar (surface tension polar component = 0), the consequent error in determining NSFE leads to an inherent error in the prediction of the contact angle of the third liquid.

## 4. Conclusions

In this research, in order to consider the surface roughness contribution to wettability, a method was developed via an extension of the BS EN 828:2013 standard which is commonly used to determine the SFE of flat surfaces. This straightforward approach was followed by the introduction of a new surface parameter, namely NSFE. The contact angles predicted by this method for liquids with a low polar surface tension component was in good agreement with the measured values, inferring the validity of the approach for low-SFE substances.

**This is the accepted manuscript (postprint) of the following article:**
M. Shaker, E. Salahinejad, *A combined criterion of surface free energy and roughness to predict the wettability of non-ideal low-energy surfaces*, Progress in Organic Coatings, 119 (2018) 123-126.
https://doi.org/10.1016/j.porgcoat.2018.02.028
[15] F.M. Fowkes, Calculation of work of adhesion by pair potential suummation, Journal of colloid and interface science, 28 (1968) 493-505.
[16] F.M. Fowkes, Donor-acceptor interactions at interfaces, The Journal of Adhesion, 4 (1972) 155-159.
[17] F.M. Fowkes, Attractive forces at interfaces, Industrial & Engineering Chemistry, 56 (1964) 40-52.
[18] D.K. Owens, R. Wendt, Estimation of the surface free energy of polymers, Journal of applied polymer science, 13 (1969) 1741-1747.
[19] C.J. Van Oss, M.K. Chaudhury, R.J. Good, Interfacial Lifshitz-van der Waals and polar interactions in macroscopic systems, Chem. Rev., 88 (1988) 927-941.
[20] C. Van Oss, R. Good, M. Chaudhury, The role of van der Waals forces and hydrogen bonds in "hydrophobic interactions" between biopolymers and low energy surfaces, Journal of colloid and Interface Science, 111 (1986) 378-390.
[21] H. Fox, W. Zisman, The spreading of liquids on low-energy surfaces. II. Modified tetrafluoroethylene polymers, Journal of Colloid Science, 7 (1952) 109-121.
[22] A.W. Neumann, R. Good, C. Hope, M. Sejpal, An equation-of-state approach to determine surface tensions of low-energy solids from contact angles, Journal of colloid and interface science, 49 (1974) 291-304.
[23] R.N. Wenzel, Surface roughness and contact angle, The Journal of Physical Chemistry, 53 (1949) 1466-1467.
[24] A. Cassie, S. Baxter, Wettability of porous surfaces, Trans. Faraday Society, 40 (1944) 546-551.
[25] BS.EN.828:2013, Adhesives. Wettability. Determination by measurement of contact angle and surface free energy of solid surface, BSI, (2013).
[26] D. Nanda, T. Swetha, P. Varshney, P. Gupta, S. Mohapatra, A. Kumar, Temperature dependent switchable superamphiphobic coating on steel alloy surface, Journal of Alloys and Compounds, 727 (2017) 1293-1301.
[27] X. Deng, L. Mammen, H.-J. Butt, D. Vollmer, Candle soot as a template for a transparent robust superamphiphobic coating, Science, 335 (2012) 67-70.
[28] B. Li, J. Zhang, Z. Gao, Q. Wei, Semitransparent superoleophobic coatings with low sliding angles for hot liquids based on silica nanotubes, Journal of Materials Chemistry A, 4 (2016) 953-960.
[29] J. Zhang, S. Seeger, Superoleophobic coatings with ultralow sliding angles based on silicone nanofilaments, Angewandte Chemie International Edition, 50 (2011) 6652-6656.
[30] J.H. Han, Innovations in Food Packaging, Elsevier, 2014.
[31] G. Mantanis, R. Young, Wetting of wood, Wood science and Technology, 31 (1997) 339-353.
[32] J. Takadoum, Materials and surface engineering in tribology, John Wiley & Sons, 2013.
[33] B.P. Binks, J.H. Clint, Solid wettability from surface energy components: relevance to Pickering emulsions, Langmuir, 18 (2002) 1270-1273.
8

**Figures**

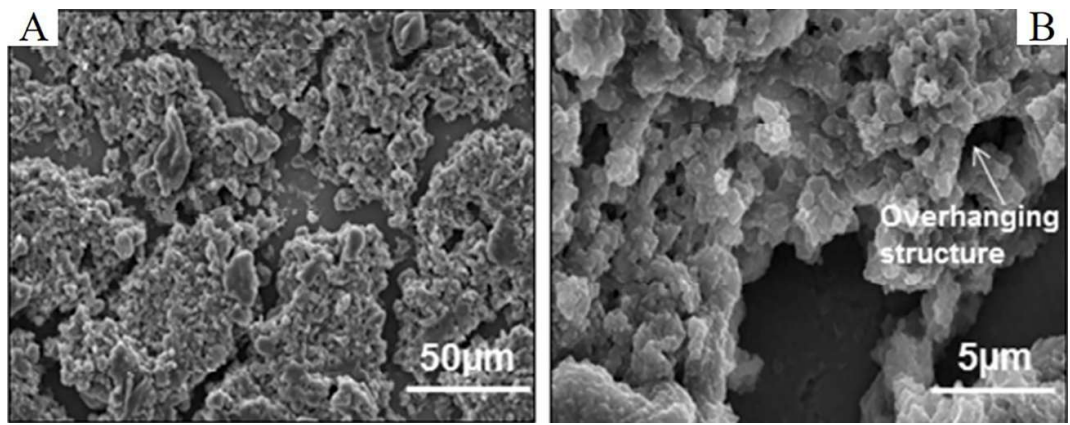

Fig. 1. Scanning electron microscopic (SEM) micrographs of the silica+PFOTS coating in two magnifications [26].

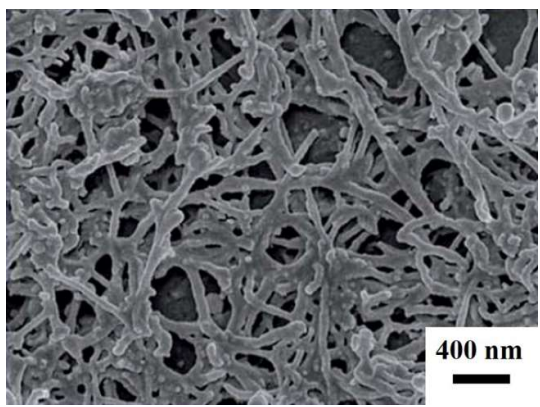

Fig. 2. SEM micrograph of the silica nanotubes coating modified with PFDTS [28].

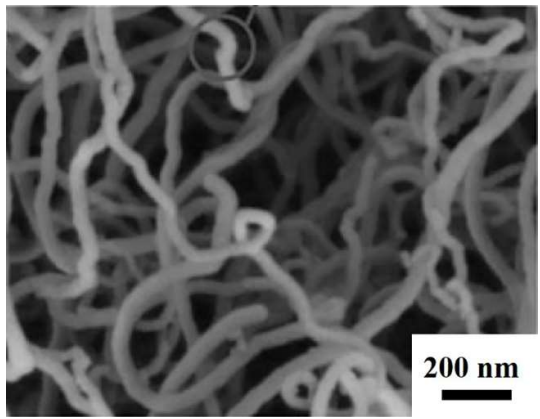

Fig. 3. SEM images of the silicone nano-filament deposit modified with PFDTS [29].

**Tables**

Table 1. Essential surface tension data of the selected liquids.

| Liquid | $\gamma_{lg}$ (mN/m) | $\gamma_{lg}^{p}$ (mN/m) | $\gamma_{lg}^{d}$ (mN/m) | Ref. |
|---|---|---|---|---|
| Water | 72.8 | 46.4 | 26.4 | [30] |
| Diiodomethane | 50.8 | 0 | 50.8 | [30] |
| Hexadecane | 28.0 | 0 | 28.0 | [31] |
| Ethylene Glycol | 48.3 | 19.0 | 29.3 | [32] |
| Toluene | 29.1 | 27.8 | 1.3 | [33] |

**This is the accepted manuscript (postprint) of the following article:**
M. Shaker, E. Salahinejad, *A combined criterion of surface free energy and roughness to predict the wettability of non-ideal low-energy surfaces*, Progress in Organic Coatings, 119 (2018) 123-126.
https://doi.org/10.1016/j.porgcoat.2018.02.028Table 2. Comparison of the contact angles measured and calculated based on the concept of NSFE. Note that the surface tension components of Liquids 1 and 2 were used to calculate NSFE. Then, the contact angle of Liquid 3 was predicted, using Eq. 2. The different precisions of the listed data is due to the different precisions reported in the used references.

| Example | Liquid 1 | Liquid 2 | NSFE (mN/m) | Liquid 3 | Predicted angle (°) | Measured angle (°) |
|---|---|---|---|---|---|---|
| I | Water | Ethylene Glycol | 0.030059±0.092751 | Hexadecane | 159±2 | 157±2.1 [26] |
| II | Water | Ethylene Glycol | 0.033272±0.012120 | Hexadecane | 162±2 | 156±1 [27] |
| II | Water | Ethylene Glycol | 0.033272±0.012120 | Diiodomethane | 157±1 | 161±1 [27] |
| II | Water | Diiodomethane | 0.044258±0.008462 | Hexadecane | 159±1 | 156±1 [27] |
| II | Water | Hexadecane | 0.052579±0.009651 | Diiodomethane | 159±1 | 161±1 [27] |
| III | Water | Diiodomethane | 0.069474±0.023617 | Hexadecane | 154.4±1 | 161.4±0.5 [29] |
| III | Water | Hexadecane | 0.035082±0.007267 | Diiodomethane | 164.2±1 | 158.6±1.6 [29] |
| III | Water | Toluene | 0.0690895±0.011812 | Hexadecane | 154.6±1 | 161.4±0.5 [29] |
| III | Water | Toluene | 0.0690895±0.0118120 | Diiodomethane | 158.1±1 | 158.6±1.6 [29] |
| IV | Water | Diiodomethane | 0.002063±0.000529 | Hexadecane | 175.3±1 | 174.4 [28] |
| IV | Water | Hexadecane | 0.001040±0.000725 | Diiodomethane | 174.0±2 | 165.7 [28] |
| IV | Water | Toluene | 0.002084±0.002316 | Diiodomethane | 171.0±2 | 165.7 [28] |
| IV | Water | Toluene | 0.002084±0.002316 | Hexadecane | 168.5±3 | 174.4 [28] |

12